\documentclass[sigconf, authorversion]{acmart}

\AtBeginDocument{%
  }

\copyrightyear{2025}
\acmYear{2025}
\setcopyright{cc}
\setcctype{by}
\acmConference[ICMI '25]{Proceedings of the 27th International Conference on Multimodal Interaction}{October 13--17, 2025}{Canberra, ACT, Australia}
\acmBooktitle{Proceedings of the 27th International Conference on Multimodal Interaction (ICMI '25), October 13--17, 2025, Canberra, ACT, Australia}
\acmDOI{10.1145/3716553.3750810}
\acmISBN{979-8-4007-1499-3/2025/10}

\usepackage{ccicons}
\usepackage{xspace}
\usepackage{url}
\usepackage{tikz}

\newcommand{\commentout}[1]{}

\usepackage{color}
\usepackage{setspace}
\definecolor{Orange}{rgb}{1,0.5,0}
\definecolor{DarkGreen}{rgb}{0,0.5,0}
\definecolor{Purple}{rgb}{0.7,0,0.7}
\definecolor{Blue}{rgb}{0.2,0.2,0.8}
\definecolor{Red}{rgb}{1.0,0.0,0.0}
\definecolor{Brown}{rgb}{0.7,0.4,0.1}

\usepackage{multicol}
\usepackage{booktabs}
\usepackage{enumitem}

\usepackage{array}
\usepackage{longtable}

\usepackage{tabularx}
\usepackage{booktabs} 
\usepackage{caption}  

\begin{document}

\title[Pinching Visuo-haptic Display]{
Pinching Visuo-haptic Display: Investigating Cross-Modal Effects of Visual Textures on Electrostatic Cloth Tactile Sensations
}

\author{Takekazu Kitagishi}
\authornote{Both authors contributed equally to this research.}
\email{kitagishi-takekazu588@g.ecc.u-tokyo.ac.jp}
\orcid{0000-0002-5370-8841}
\affiliation{%
  \institution{The University of Tokyo}
  \institution{ZOZO Research}
  \city{Tokyo}
  \country{Japan}
}

\author{Chun-Wei Ooi}
\authornotemark[1]
\email{ooic@g.ecc.u-tokyo.ac.jp}
\orcid{0000-0002-1517-8755}
\affiliation{%
  \institution{The University of Tokyo}
  \city{Tokyo}
  \country{Japan}
}

\author{Yuichi Hiroi}
\email{y.hiroi@cluster.mu}
\orcid{0000-0001-8567-6947}
\affiliation{%
  \institution{Cluster Metaverse Lab}
  \city{Tokyo}
  \country{Japan}
}

\author{Jun Rekimoto}
\email{rekimoto@acm.org}
\orcid{0000-0002-3629-2514}
\affiliation{%
  \institution{The University of Tokyo}
  \institution{Sony CSL Kyoto}
  \city{Tokyo/Kyoto}
  \country{Japan}
}

\renewcommand{\shortauthors}{T. Kitagishi, C. Ooi, Y. Hiroi, and J. Rekimoto}

\begin{abstract}
    Tactile representations of fabrics are critical in virtual retail and VR fitting applications. A pinching tactile display (PTD) using electrostatic adhesion is a fabric-type display that can flexibly adjust tactile sensations by adjusting voltage and frequency, and is expected to be applied to the remote transmission of fabric tactile sensations. However, no significant differences in tactile sensation other than roughness were observed. Building on established research in cross-modal perception, we investigated whether overlaying visual fabric textures could alter the tactile expressiveness of PTDs. We developed a system that integrated a systematically controlled conductive fabric-based PTD with spatially registered VR fabric simulations. Participants rated four tactile properties (roughness, stiffness, thickness, warmth) while touching the same conductive substrate under different visual texture conditions (denim, gauze, toweling, voile). The results showed that voile texture significantly affected stiffness (p=.00887); toweling texture tended to affect warmth (p=.0658); and denim texture tended to affect roughness (p=.0972). These results suggest that virtual texture overlay can affect the warmth, stiffness, and roughness of the PTD. Since the PTD can only change roughness, these results suggest that texture overlay enables the display of tactile sensations that the PTD alone cannot express. On the other hand, no significant influence of texture on thickness was observed. This suggests the pinching motion provided a highly reliable tactile cue, which may have led to visual information being underrated and thus not reaching a significant level. These findings offer strategic insights for employing visual textures to enhance perceived fabric properties, paving the way for deployment in virtual fashion retail and VR fitting applications.
\end{abstract}

\ccsdesc[300]{Hardware~Sensors and actuators}
\ccsdesc[300]{Hardware~Haptic devices}
\ccsdesc[100]{Hardware~Electro-mechanical devices}
\ccsdesc[100]{Computer systems organization~Sensors and actuators}
\ccsdesc[500]{Human-centered computing~Haptic devices}
\ccsdesc[300]{Human-centered computing~Mixed / augmented reality}
\ccsdesc[300]{Human-centered computing~Virtual reality}
\ccsdesc[500]{Hardware~Emerging interfaces}

\keywords{MultiModal; Cross-modal Effect; VR; Physical Simulation; Haptic Display; }

\begin{teaserfigure}
  \includegraphics[width=\textwidth]{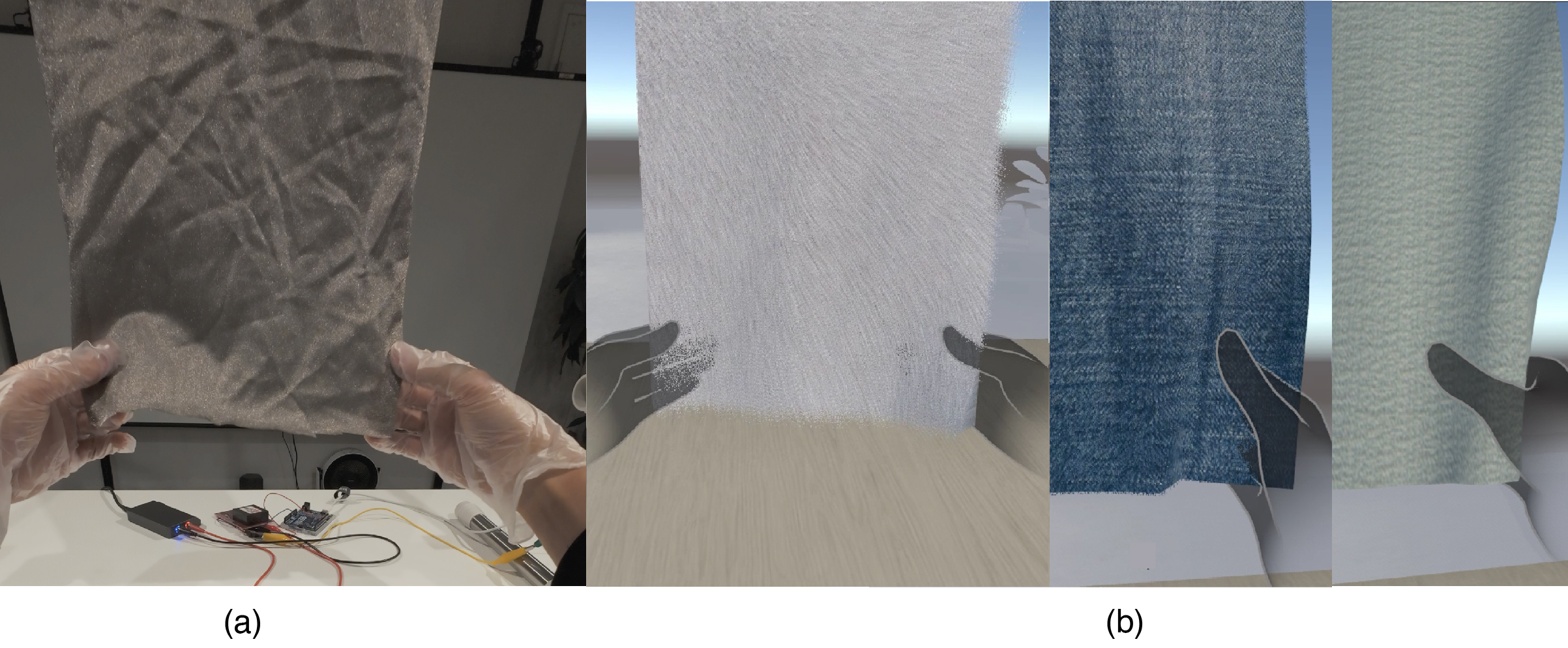}
  \caption{
    Our setup to investigate cross-modal effects: We explored how visual textures influence tactile perceptions of cloth. (a) First person view of a participant interacting with a Pinching Tactile Display (PTD)~\cite{PTD}, which provides tactile feedback via programmable electrostatic adsorption. 
    (b) In VR conditions, a virtual cloth with modifiable texture renderings (toweling, denim, gauze) is placed in the same position as the physical PTD using a tracking controller. Fur and bumps are applied in various degree for different fabric type to mimic their physical counterpart.  
  } 
    \Description[Teaser figure. Physical and virtual reality components of the pinching visuo-haptic display system]{Two-panel setup: 
    (a) Hands in gloves touching conductive cloth connected to a tactile driver electronics;
    (b) A VR view with virtual hands reaching toward fabric samples: toweling, denim, gauze.}
  \label{fig:teaser}
\end{teaserfigure}

\received{2 May 2025}
\received[revised]{12 June 2025}
\received[accepted]{7 July 2025}

\maketitle

\section{Introduction}\label{sec:introduction} 
\begin{figure}[t]
      \centering
      \includegraphics[width=0.8\linewidth]{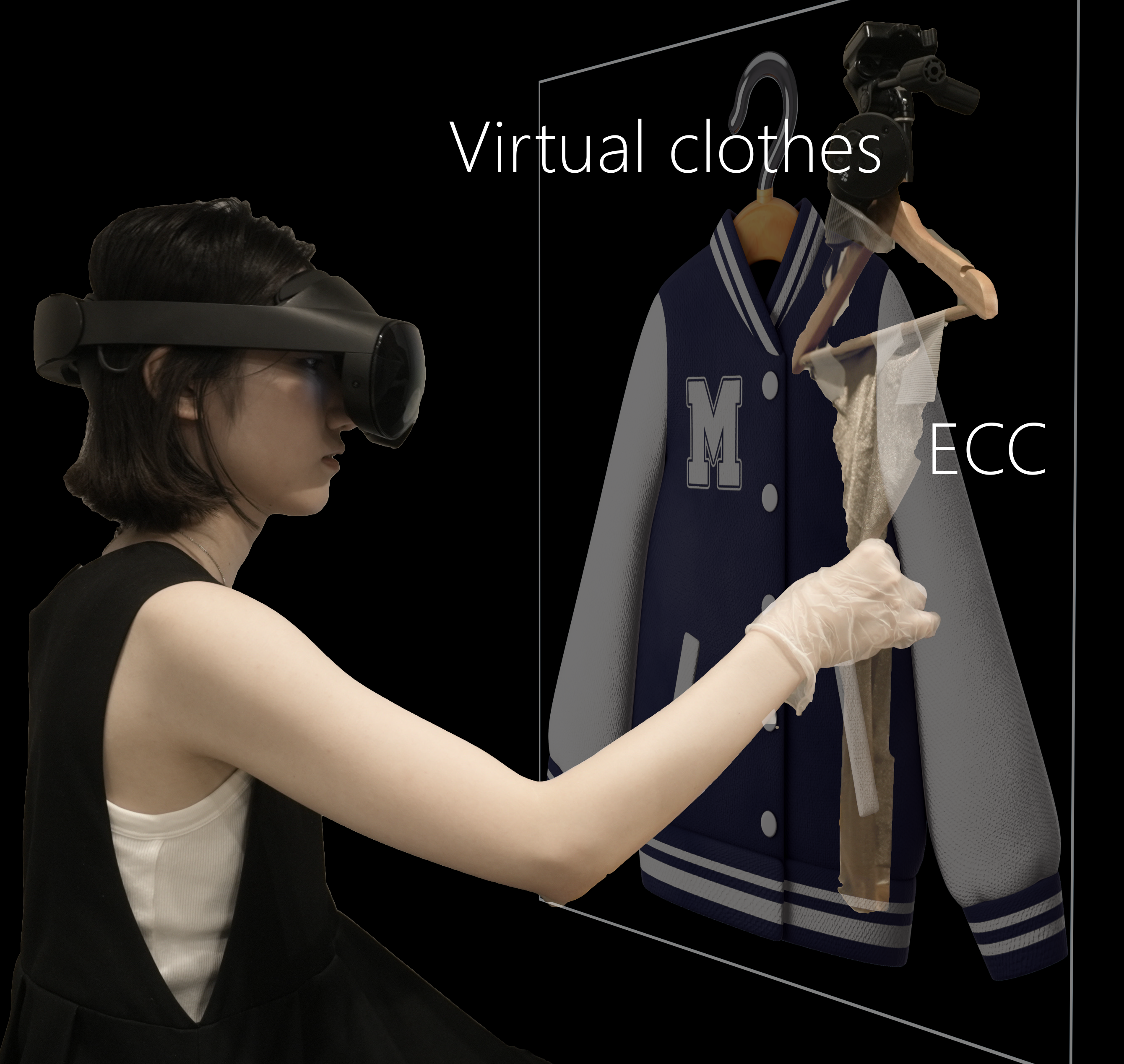}
      \caption{
      Example of an online shopper interacts tangibly with the fabric of the virtual clothes in the VR with the help of ECC in the real world.
      }  
      \Description[User wearing VR headset interacting with virtual clothing while touching physical conductive cloth]{A person wearing a VR headset reaches toward a blue varsity jacket while touching a piece of conductive fabric labeled "ECC." This demonstrates a virtual fashion application with haptic feedback.}
      \label{fig:shop}
    \end{figure}

Tactile representations of fabrics are critical in virtual retail~\cite{retail,VRshop} and VR fitting applications~\cite{shoppingVR}, where users evaluate the authenticity, quality, and comfort of clothing. When buying clothes in physical stores, most people don't just look at the design; they also touch the fabric to see if it feels the way they like. At this time, humans evaluate the tactile properties of fabrics by stroking and pinching~\cite{atkinson2013tactile, Yu2016Exploring, cary2013exploring}. If VR environments incorporate similar physical feedback and gesture capabilities, they can recreate the common real-world act of checking the feel of fabrics in a VR space, enabling users to "shop" as if in a physical store, as shown in Figure \ref{fig:shop}. 
However, accurately reproducing the flexibility and subtle textural nuances of fabrics while preserving these natural interaction modalities remains challenging.


Traditional fabric tactile displays rely on limited fabric patterns and only allow users to trace their surfaces~\cite{HapticRevolver, degraen2020envisioning, kitagishi2023telextiles}. In contrast, the Pinching Tactile Display (PTD)~\cite{PTD} enables gestures that more closely resemble real-world fabric interactions. This system dynamically changes tactile sensations by applying high-voltage pulse waves to conductive fabric, allowing users to explore tactile sensations by pinching and rubbing as they would with normal fabric. However, PTD is limited to generating tactile sensations similar to the base conductive fabric. It remains challenging to replicate tactile characteristics beyond those of the original material. Additionally, no significant differences in tactile sensation were observed other than in roughness~\cite{PTD}.

A promising approach to overcome this limitation is the visuo-haptic cross-modal effect, where visual information influences tactile perception~\cite{840369, ernst2002humans,SUN2016353,yanagisawa2015effects}. Studies have shown that when visual and tactile signals conflict during sensory integration, vision often dominates~\cite{doi:10.1126/science.143.3606.594, 1492747, gibson1933adaptation}. Exploiting this phenomenon could potentially provide tactile sensations beyond what the PTD alone can provide by presenting different visual textures. 

In fabric tactile perception, four fundamental properties have been identified : roughness, stiffness, thickness, and warmth~\cite{atkinson2013tactile, behery2005effect, fan2004clothing, li2024investigation}. 
Previous research has investigated these tactile sensations as separate properties~\cite{li2024investigation}. All of these properties can be influenced by visual cues. For example, thickness perception can be influenced by visually altering an object's apparent size~\cite{abtahi2018visuo, zhang2021using}, and warmth perception shifts when the visual color of a fabric changes~\cite{atav2024investigation}.
However, these experiments use simplified interactions or static materials. Interaction variables, such as movement trajectory and gesture type, have been reported to affect the strength of the illusion~\cite{10.1145/3491102.3517671}. Since pinching and rubbing are more complex, these gestures can disrupt visual illusions. In addition, fabrics deform dynamically when touched, which can affect the cross-modal effect~\cite{punpongsanon2020flexeen}.
Therefore, our research investigates how viewing different virtual fabric textures affects tactile perception when users perform gestures that resemble real-world interactions with fabrics.

To investigate this question, we developed Pinching Visuo-haptic Display, which presents a physically simulated virtual fabric within a VR space, at the location of the physical conductive fabric. When users touch the conductive fabric, the virtual fabric moves synchronously, creating a coherent cross-modal experience. Users can naturally explore fabric sensations in a VR space, through gestures typically used in fabric evaluation, such as pinching and stroking, as shown in Figure~\ref{fig:interaction}.

Using this system, we conducted user experiments to examine how visual textures modulate the tactile perception of conductive fabrics. Participants evaluated the same conductive cloth under four different visual conditions: denim, gauze, toweling, and voile (Figure~\ref{fig:teaser}). The results mainly showed that the voile texture significantly affected perceived stiffness ($p = .00887$), and the toweling texture tended to affect perceived warmth ($p = .0658$). These results suggest that visual overlays can convey sensations of stiffness and warmth that the physical device alone cannot.


Our main contributions are as follows:
\begin{itemize}[leftmargin=*]
\setlength{\parskip}{0cm}
    \item We introduced a Pinching Visuo-haptic Display that spatially synchronizes a physically simulated virtual fabric in VR with a conductive-cloth tactile display. This display supports natural gestures for interacting with cloth, such as pinching and stroking.
    \item We demonstrate that visual textures influence tactile perception. For example, denim evokes the sensation of roughness, voile evokes the sensation of stiffness, and toweling evokes the sensation of warmth. Meanwhile, thickness remains visually robust under pinching.
    \item Since the PTD primarily affects roughness, combining it with visual overlays effectively expands the display's range to include stiffness and warmth. This demonstrates how visual augmentation can compensate for hardware limitations.
\end{itemize}

\section{Related Work}\label{sec:related}
\subsection{Visuo-Haptic Interaction with Cloth}
Research on visual-haptic interactions has demonstrated that tactile perception can be significantly influenced by visual cues~\cite{840369,ernst2002humans, doi:10.1126/science.143.3606.594, 1492747, gibson1933adaptation}. When examining cloth tactile perception specifically, prior studies suggest that different tactile properties should be analyzed individually rather than as a unified whole~\cite{li2024investigation}. 
According to the literature~\cite{soufflet2004comparison, behery2005effect, fan2004clothing, ciesielska2012hand}, tactile properties of fabrics that influence handle can be categorized as follows:
\begin{itemize}[
    leftmargin=*,    
    topsep=0pt,      
    partopsep=0pt,   
    parsep=0pt,      
    itemsep=0pt      
]
  \item Surface properties: compression properties, friction, surface irregularity, related to roughness (rough–smooth).
  \item Physical properties: thickness, mass per unit area, related to perceptions of thickness (thick–thin).
  \item Mechanical properties: extensibility, bending properties, shear, related to stiffness (stiff–flexible).
  \item Thermal properties: conductivity, related to warmth (warm–cool).
\end{itemize}
Furthermore, previous studies indicate that each of these properties can potentially be influenced by visual information.

For roughness perception, studies have demonstrated that visual appearance can significantly alter how a surface feels when touched~\cite{klatzky1999tactile, ramalho2013friction, Guenther2022smooth}. In particular, S. Guenther et al.~\cite{Guenther2022smooth} investigated how visual appearance influences perceived roughness when touching textures in VR, which closely relates to our study. We aim to confirm whether changes in roughness perception also occur in the context of fabric interactions such as pinching.

Stiffness~\cite{wolf2015illusion, punpongsanon2020flexeen, weiss2023using, cellini2013visual, ban2019displaying, drewing2009haptic, li2016evaluation}, and thickness~\cite{10.5555/933178, abtahi2018visuo, zhang2021using} perceptions have also shown susceptibility to visual influence. P. Punpongsanon et al.~\cite{punpongsanon2020flexeen} demonstrated that visually manipulated deformation characteristics could alter perceived stiffness of virtual objects, while Abtahi et al.~\cite{abtahi2018visuo} showed that altering an object's apparent size visually influenced the perception of its physical size. These studies typically used virtual objects or rigid surfaces rather than fabric specifically, leaving a gap in understanding how these effects manifest with textiles.

Warmth perception has been explored in relation to color and texture visualization, with research indicating that warmer colors can influence thermal perception~\cite{atav2024investigation}. However, these studies were often conducted with static images rather than in interactive VR environments.

A significant limitation of existing research is that many visuo-haptic experiments use simplified interactions or static materials. As noted by M.Kurzweg et. al.~\cite{Kurzweg2024survey}, visual-haptic illusions can be less effective when users receive strong contradicting tactile signals. Complex exploratory gestures—such as those commonly used when evaluating fabrics—might disrupt visual illusions, as interaction variables like movement trajectory and grasping type can affect illusion strength~\cite{10.1145/3491102.3517671}. Our research addresses this gap by studying visual texture effects in a system that supports common fabric interaction gestures.

\begin{figure}[t]
 \centering
 \includegraphics[width=\linewidth]{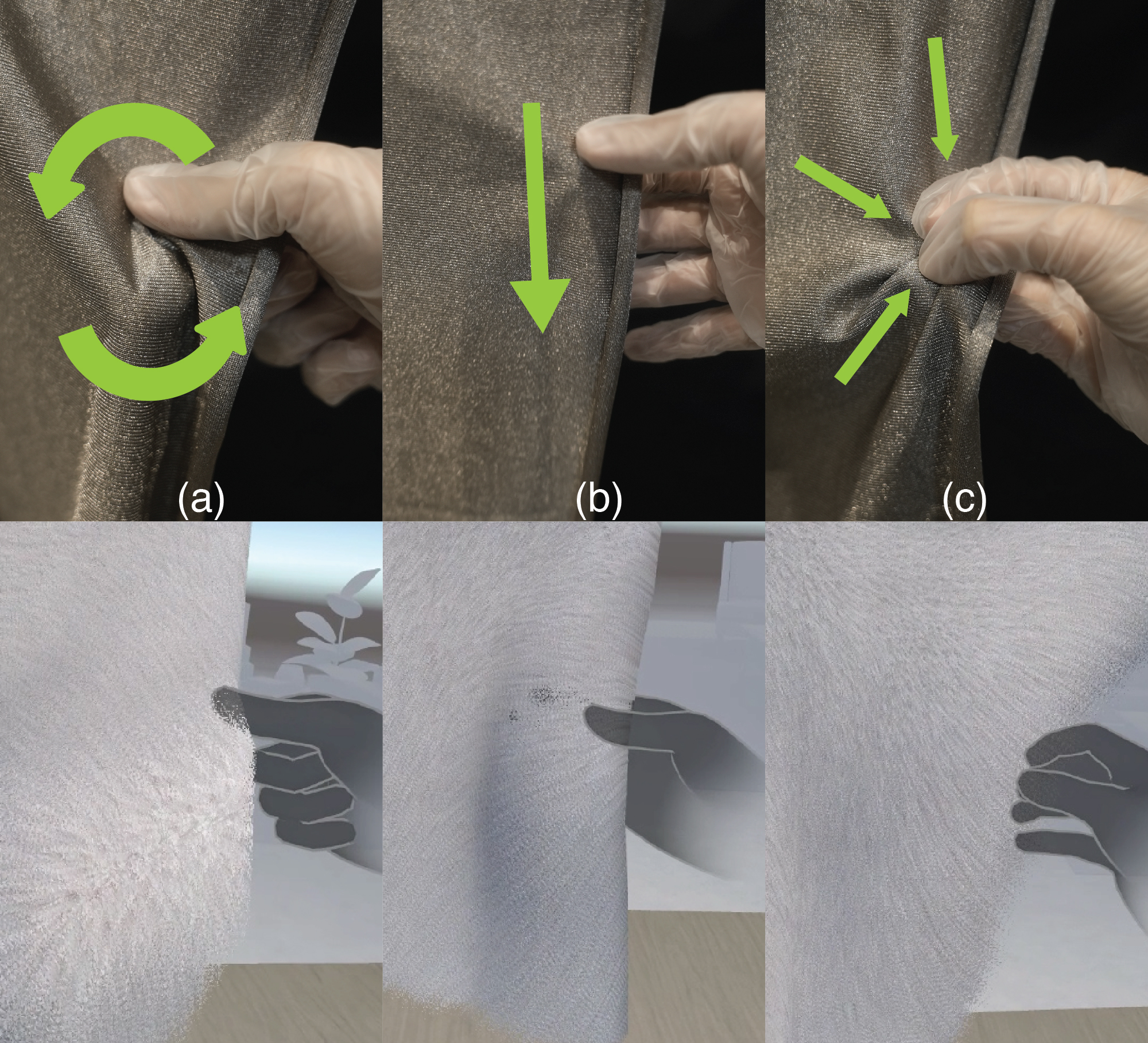}
 \caption{Top-The type of interaction gesture and its respective direction of motion shown in green arrow: (a)pinching (b)stroking (c)scrunching. Bottom-The corresponding gesture in VR.}  
 \label{fig:interaction}
    \Description[Hand interaction gestures with cloth]{Six panels showing hand gestures on conductive cloth (top) and VR visualization (bottom): (a) Pinching with a circular motion; (b) Stroking downward; and (c) Scrunching with diagonal movement.}
\end{figure}

\subsection{Fabric-Based Tactile Displays}
Creating realistic fabric tactile sensations in VR presents unique challenges due to the soft, flexible nature of textiles and the subtle differences in texture that humans readily perceive~\cite{Yu2016Exploring, Xing2024Investigating}. Previous research has developed various approaches to fabric tactile feedback, each with distinct limitations.

Touchscreen-based systems have been developed to simulate fabric textures through vibration or electrostatic friction~\cite{KatumuGorlewicz2016, KimOsgoueiChoi2017}. While these approaches offer precise control, they face challenges with flat, stiff surfaces, limiting interaction to surface tracing. This results in a lack of the feel of soft, flexible, stretchable fabrics and does not support common fabric exploration gestures like pinching or scrunching. The iShoogle system~\cite{atkinson2013tactile, orzechowski2016pinching} has attempted to address this issue by developing specialized interfaces for textile exploration, identifying three key gestures for fabric evaluation: pinch, stroke, and scrunch~\cite{atkinson2013tactile}. However, these systems still rely on flat, stiff surfaces.

Alternative approaches use specifically engineered materials, such as 3D printed hair structures~\cite{DegraenZennerKruger2019} or pneumatic tactile displays~\cite{Cai2024ViboPneumo, YaoPneutouch2025}. While these methods can provide interesting tactile sensations, they still struggle to replicate the nuanced properties of actual fabrics since they do not utilize real fabric materials.

Several researchers have developed systems that showcase multiple physical fabric samples, such as HapticRevolver~\cite{HapticRevolver}, Haptic Palette~\cite{10.1145/3393914.3395870}, and Telextiles~\cite{kitagishi2023telextiles}. While these approaches offer authentic fabric sensations, they are limited by the number of different fabric samples they can physically incorporate, restricting their versatility in VR applications.


A promising approach uses electrically conductive cloth (ECC) to create dynamic tactile displays~\cite{PTD}. The Pinching Tactile Display (PTD) uses electrostatic adhesion to modulate the tactile sensation of conductive fabric. This method preserves the natural softness and flexibility of fabric while allowing dynamic adjustment of its tactile properties through voltage and frequency modulation. However, while the PTD can increase perceived roughness by adjusting pulse wave voltage, it has not been shown to significantly affect stiffness, thickness, or warmth on its own. Additionally, these systems are limited in their ability to produce sensations that differ from those of the base conductive cloth, which presents a challenge when attempting to simulate tactile characteristics not originally present in the base material.

Our work builds upon fabric-based tactile displays, specifically the Pinching Tactile Display approach, while addressing its expressiveness limitations through visual augmentation rather than additional physical mechanisms.

\subsection{Texture Rendering on Cloth}

Realistic visualization of fabric textures in virtual environments presents significant technical challenges. Previous approaches to fabric visualization in VR and AR can be categorized into several methods.

Screen-based visualization approaches have been explored extensively in virtual try-on applications~\cite{ehara2006texture, hauswiesner2013virtual, cao2015educational, saakes2016mirror}. These systems overlay textile textures onto 2D images or 3D models viewed on monitors. While they can create visually convincing representations, they suffer from limited immersion due to the constrained viewing angle and lack of direct interaction possibilities.

Projection mapping techniques have been used to overlay textures onto physical objects, including fabrics. FleXeen~\cite{punpongsanon2020flexeen} demonstrated that projection mapping could alter perceived fabric stiffness through visual manipulation. These approaches allow users to view augmented fabrics directly with their eyes, potentially reducing sensory conflicts. However, they typically require controlled lighting environments and precise calibration, limiting their flexibility.

Augmented reality approaches using see-through displays~\cite{bau2012revel} or video pass-through~\cite{Inami2000} have been explored for texture visualization. By detecting fabric regions and overlaying virtual textures, these systems can create the illusion of different fabrics. However, video see-through methods create two modalities (real and virtual), which result in discomfort. 

For VR-based fabric rendering, Y.Tsukuda and M.Koeda~\cite{tsukuda2019difference} not only created 3DCG fabric by physical simulation but also evaluated the touch of the cloth in a VR space. VR-based fabric rendering offers the advantage of complete control over the visual environment, which result in immersion and satisfaction. Using physical simulation enables versatility and convenience. The cloth in VR space can be easily replaced, enabling immersive experiences in mobile environments without special experimental equipment. However, the touch of the cloth was created by vibration feedback from VR controllers. This work investigates how virtual texture overlays on physical conductive cloth affect tactile perception during natural fabric interaction gestures in VR.

\section{System Configuration}

\begin{figure}[tb]
 \centering
 \includegraphics[width=0.9\linewidth]{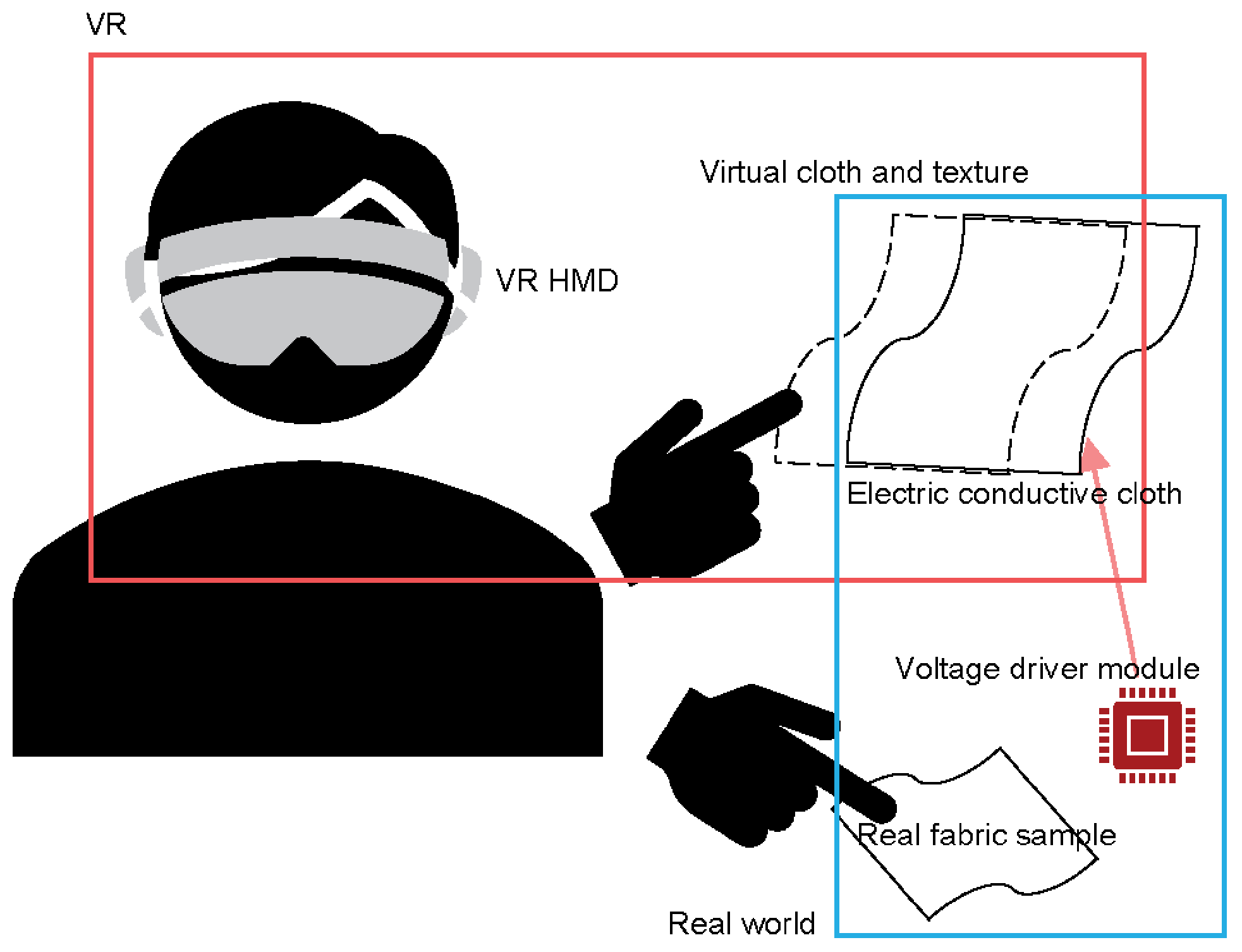}
 \caption{Schematic diagram of the pinching visuo-haptic display. The red rectangle represents the VR modality, while the blue rectangle represents the real world. In VR, the fabric visuals appear at the same position as the ECC in the real world. The ECC is driven by a voltage driver module regulated by an Arduino. A real fabric panel is also prepared for comparisons.}  
 \Description[System architecture diagram]{System Architecture: The VR domain (left, red border) shows the user wearing a headset, while the real-world domain (right, blue border) shows the conductive cloth and voltage driver for tactile feedback.}
 \label{fig:system-overview}
\end{figure}

Figure~\ref{fig:system-overview} shows a schematic of our proposed visuo-haptic display, consisting of two key modules: a VR modality and a PTD that provides electrotactile feedback.
We suspended a conductive cloth connected to a PTD tactile driver and rendered its virtual counterpart at the exact corresponding position in VR. The virtual fabric's movements synchronize with the physical conductive cloth through spatial registration. The system supports common fabric exploration gestures, such as pinching and stroking, allowing users to experience visuo-haptic fabric sensations with and without a VR headset to evaluate cross-modal effects.
This spatial registration allows users to simultaneously see the virtual fabric while physically touching the electrically conductive cloth, creating a cohesive visual and tactile experience.

\subsection{VR Modality}
\label{sec:vr-modality}
To achieve precise spatial alignment between the physical and virtual fabrics, we employed real-time tracking to ensure that any movement or manipulation of the physical fabric is mirrored in the virtual environment. This setup enables realistic user interactions by accurately reflecting fabric behaviors, ensuring the integrity of tactile feedback with corresponding visual stimuli.

    \begin{figure}[t]
      \centering
      \includegraphics[width=1\linewidth]{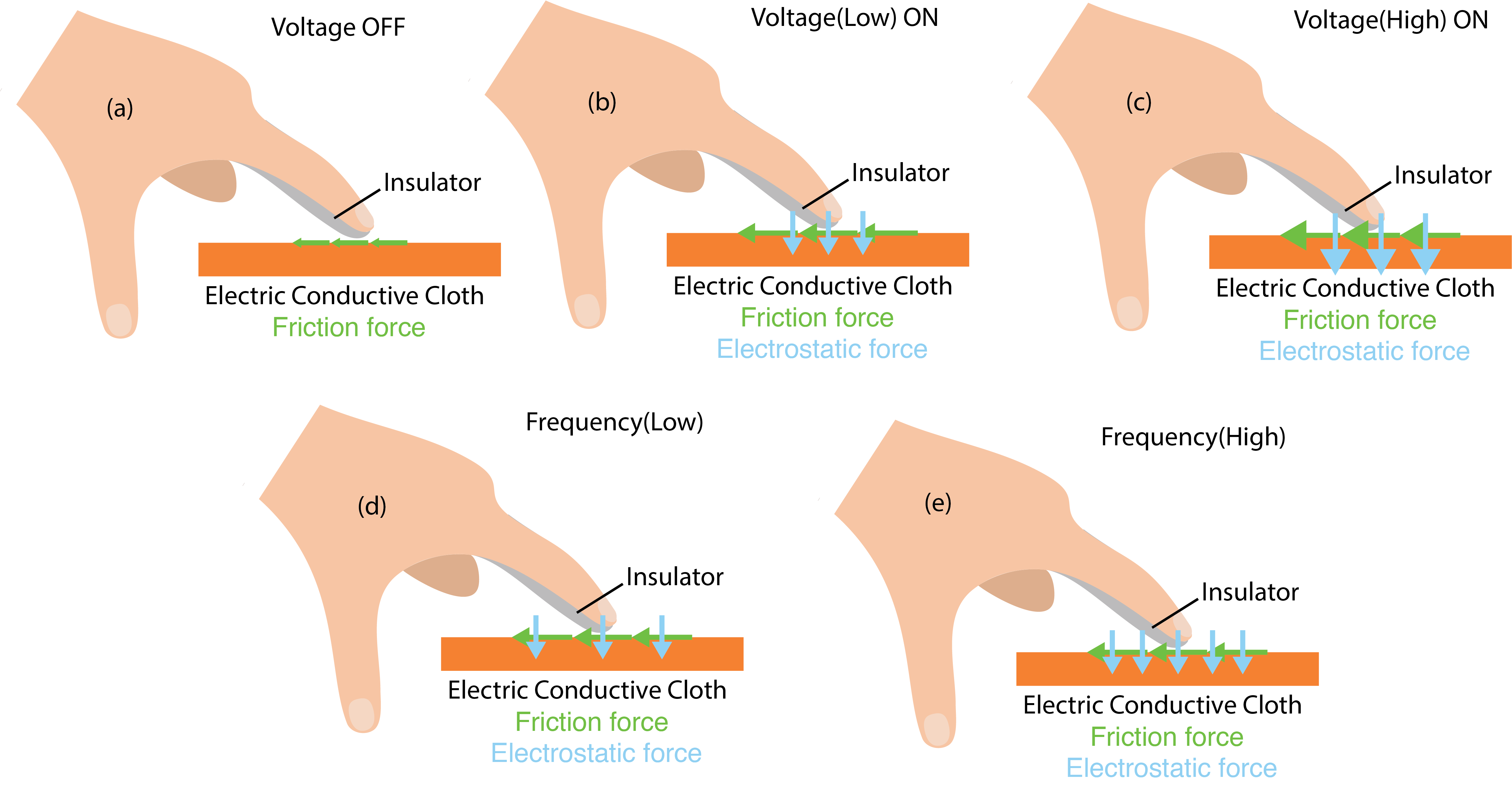}
      \caption{
      Electrostatic tactile modulation in PTD. 
      Voltage applied to conductive cloth generates electrostatic forces that induce periodic skin deformation through varying frictional forces.  Tactile perception varies as parameters are adjusted: (a) zero voltage produces negligible friction, while (b) low and (c) high voltages correspond to proportional increases in perceived frictional force. Frequency modulation independently affects texture perception - (d) low frequencies (50 Hz) produce a sense of roughness, while (e) high frequencies (200 Hz) produce a sense of finer texture.
      }  
      \Description[Electrostatic force mechanism]{Five panels showing fingers touching conductive cloth:
        (a) Voltage off with friction forces
        (b-c) Low/high voltage adding electrostatic forces
        (d-e) Low/high frequency variations affecting tactile perception}
      \label{fig:mechanism_of_electrovibration}
    \end{figure}

\subsection{Pinching Tactile Display}
\label{sec:tactile-feedback-modality}
For the tactile component, we use the PTD. As shown in Fig.~\ref{fig:mechanism_of_electrovibration}, when a voltage is applied to the conductive fabric, electrostatic forces are generated between the fabric and the user's finger, creating periodically varying frictional forces that induce dynamic tactile sensations. At low voltages (100 V), users perceive minimal frictional force; at high voltages (300 V), they experience greater frictional feedback. 

Users wear ultra-thin, non-conductive gloves that serve the dual purpose of insulating the fingers from the electric current while maintaining sufficient tactile sensitivity. The system generates different tactile sensations by adjusting both the power supply voltage and the transistor switching frequency, creating a range of perceptual effects while maintaining the inherent thinness and flexibility of the fabric. This approach allows users to interact with the fabric using natural exploratory movements while experiencing dynamically variable tactile feedback.

    \begin{figure}
     \centering
     \includegraphics[width=\linewidth]{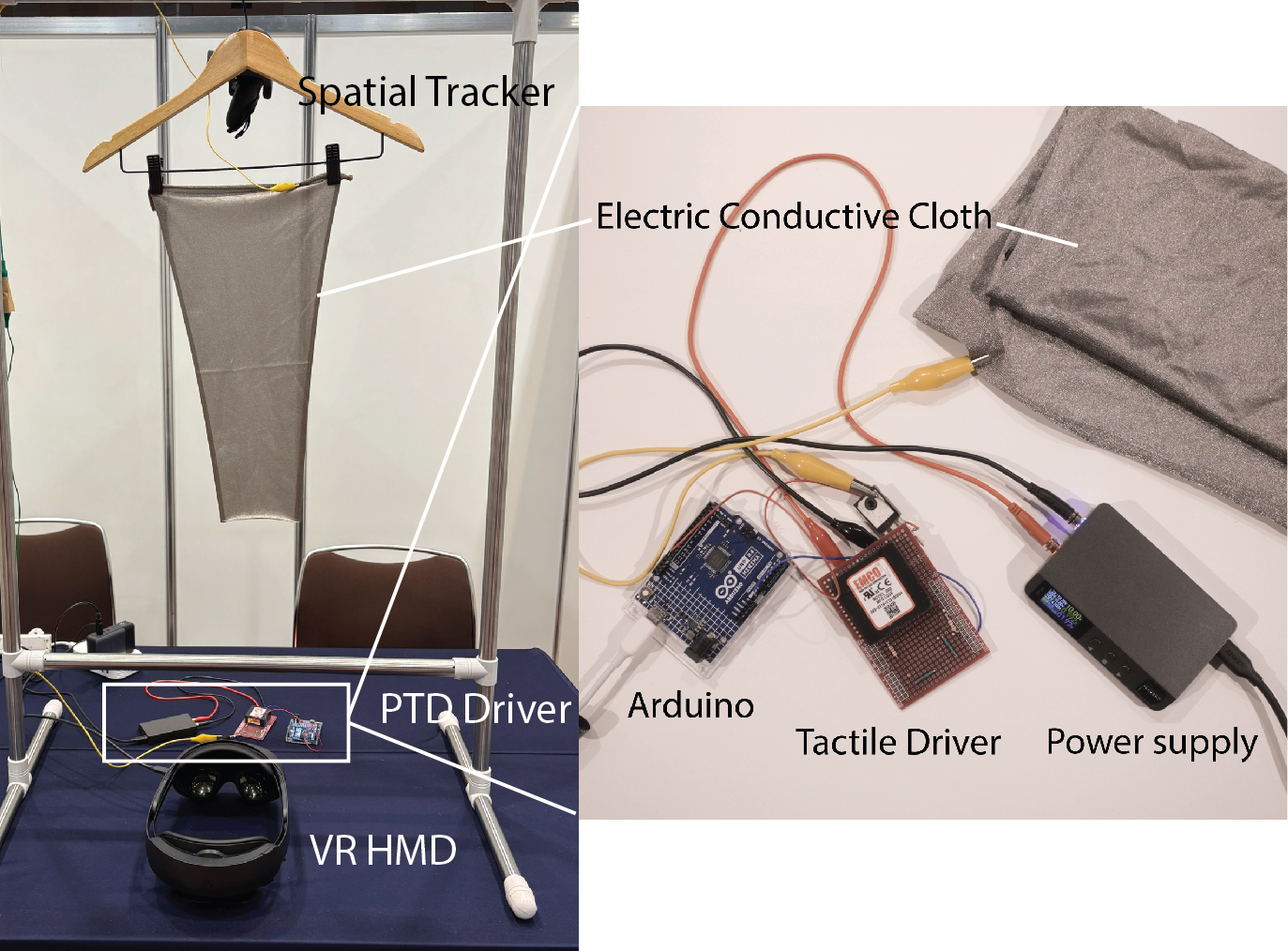}
     \caption{Left: Our proposed visuo-haptic setup that involves the PTD driver, a VR HMD and a ECC which is being hung using a hanger with a spatial tracker attached to it. Right: A closeup shot of the PTD driver consisting of an Arduino, Tactile Driver, and power supply.} 
     \Description[Hardware setup]{The left shows a cloth suspended from a hanger with a spatial tracker and a VR headset. The right shows an Arduino, a tactile driver, and a power supply connected to a conductive cloth.}
     \label{fig:setup}
    \end{figure}

\subsection{Implementation}
\label{sec:implementation}
In this study, we investigated the overall cross-modal effects of fabric interactions within a VR environment, especially in the context of VR shopping.  


\subsubsection{VR Setup}
Figure~\ref{fig:setup} shows the overall setup.
We used a Meta Quest Pro VR headset~\cite{meta2022questpro}, employing the Unity~\cite{unity2025engine} game engine for simulation and rendering.

    \begin{table}[h]
        \caption{Fabric properties and their corresponding stiffness and damping coefficients used in our cloth simulation.}         
        \centering
        \footnotesize
        \begin{tabular}{@{}lcccc@{}}
            \toprule
            \begin{tabular}[c]{@{}l@{}}\textbf{Type of}\\\textbf{Fabric}\end{tabular} & 
            \begin{tabular}[c]{@{}c@{}}\textbf{Spring (Stretch)}\end{tabular} & 
            \begin{tabular}[c]{@{}c@{}} \textbf{Spring (Bend)}\end{tabular} & 
            \begin{tabular}[c]{@{}c@{}} \textbf{Damping} \end{tabular} \\
            \midrule
            Denim       & Low & Low  & High \\
            Towelling   & High  & High  & High \\
            Gauze   & High & Medium   & Low \\
            Voile       & Medium  & High    & Low \\
            \bottomrule
            *Low:0.1 ; Medium:0.5 ; High:0.9 
        \end{tabular}
        \Description[Fabric simulation parameters]{Spring and damping coefficients for four fabrics: Denim (Low/Low/High), Towelling (High/High/High), Gauze (High/Medium/Low), Voile (Medium/High/Low). Scale: Low=0.1, Medium=0.3, High=0.9.}
        \label{tab:fabric_properties}
    \end{table}

\paragraph{Cloth Simulation}
To simulate cloth behavior accurately, we implemented a particle network spring mass system~\cite{clothsimulation} in Unity. 
Each particle of the virtual cloth is represented by a mesh of mass points (particles) connected by springs that enforce stretchiness, bendiness and damping constraints. Spring(stretch) is the properties that affect the elasticity of the fabric when pulled. Spring(bend) is the properties that affect the stiffness of the fabric when bent. Damping coefficient is the properties that influence the how quickly the fabric settles into a static state after the force is applied, high damping coefficient indicates heavier fabric. The force of each spring follows Hooke’s Law:

\begin{equation}
   \mathbf{F}_{\text{spring(stretch,bend)}} = -k \left(\mathbf{p}_i - \mathbf{p}_j - \mathbf{L}_0\right),
\end{equation} 

where k is the spring coefficient, \(\mathbf{p}_i\) and \(\mathbf{p}_j\) are the positions of the connected mass points, and \(\mathbf{L}_0\) is the rest length vector of the spring. A damping force is also introduced to add decay to the simulation:

\begin{equation}
    \mathbf{F}_{\text{damping}} = -c \left(\mathbf{v}_i - \mathbf{v}_j\right),
\end{equation}

where \(c\) is the damping coefficient, and \(\mathbf{v}_i, \mathbf{v}_j\) are the velocities of the connected mass points. 

For cloth simulation, we assigned 3 normalized values[0,1] to each of the parameters; 0.1(low), 0.5(medium), 0.9(high), as we empirically tuned the spring and damping coefficient values of the virtual cloth based on their real counterparts (denim, gauze, toweling, and voile), as shown in Table~\ref{tab:fabric_properties}. 
 Denim fabric is rigid, stiff and heavy, so we assigned low to spring(stretch), low to spring(bend) and high to damping coefficient. Towelling fabric is elastic, flexible and heavy, so we assigned high to spring (stretch), high to spring (bend) and high to damping coefficient. Gauze fabric is rigid, somewhat stiff and light, high to spring (stretch), medium to spring (bend) and low to damping coefficient are assigned. Voile fabric is somewhat rigid, flexible, and light, medium to spring (stretch), high to spring (bend) and low to damping coefficient are assigned.



    \begin{figure}[tb]
     \centering
     \includegraphics[width=0.6\linewidth]{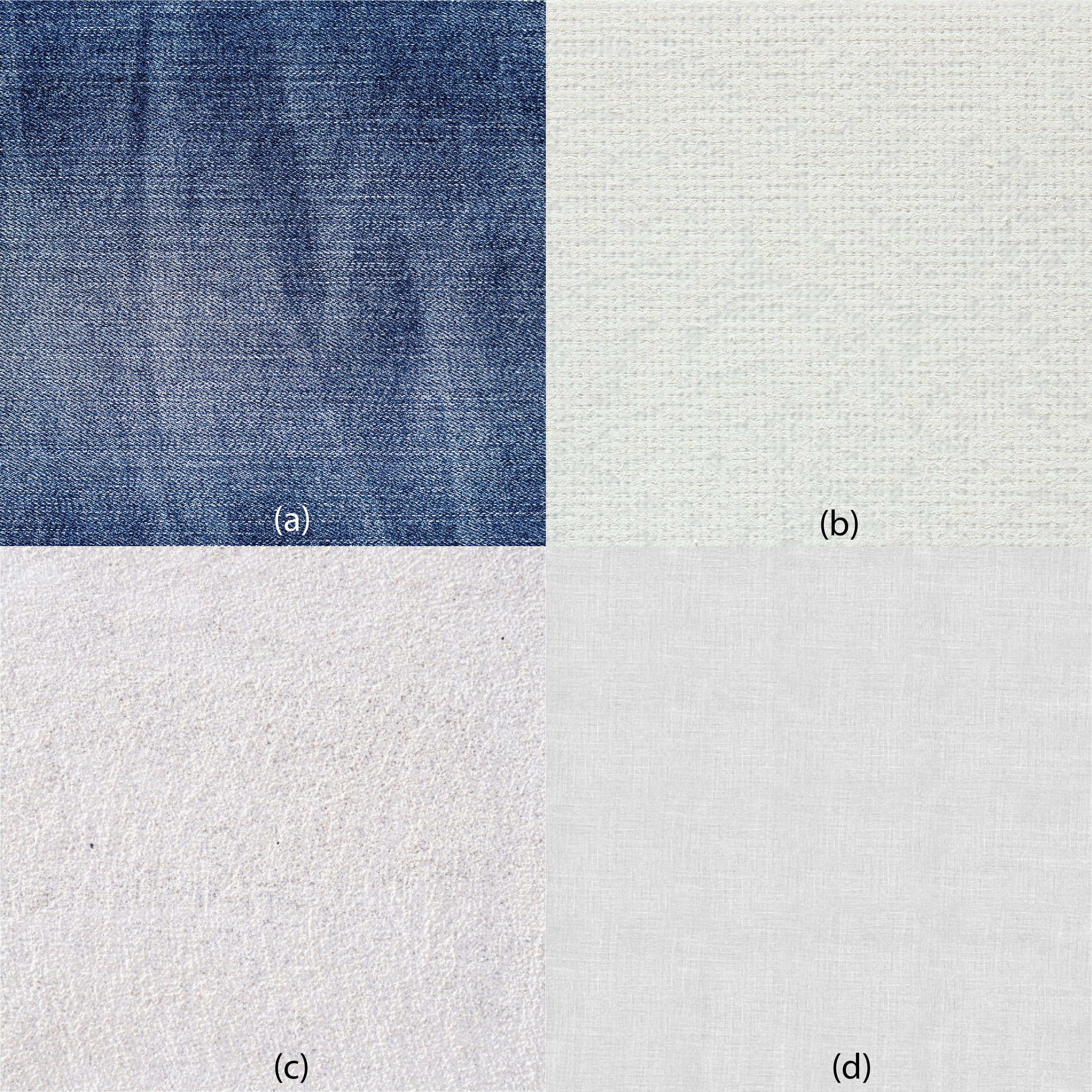}
     \caption{Textures used in the virtual cloth. (a) denim, (b) gauze, (c) towelling, and (d) voile.} 
     \Description[Four fabric texture samples]{Four fabric textures used in VR experiments: (a) blue denim with a diagonal twill pattern, (b) gauze with an open weave, (c) terry cloth with fuzzy loops, and (d) smooth voile fabric.}
     \label{fig:textures}
    \end{figure}

\paragraph{Texturization of the Virtual Cloth}
Figure~\ref{fig:textures} shows the texture.
We applied high-resolution texture scans of Towel, Jeans, Gauze, and Voile from Quixel Megascan~\cite{quixel2025megascans}. Additionally, we employed a fur shader~\cite{stratasoft2025furfx} to enhance perceived surface roughness. Shader implementation balanced visual appearance and performance optimization for VR deployment.

\paragraph{Interaction with the Virtual Cloth}
To facilitate intuitive interactions, we implemented Meta SDK hand tracking. We converted each particle of the virtual cloth into an interactable grab point, enabling precise "pinching" actions. Capsule colliders attached to hand joints further enhanced the realism of "stroking" and "scrunching" interactions, as shown in Figure~\ref{fig:interaction}.

\paragraph{Spatial Positioning of the Virtual Cloth}
For accurate alignment, the Meta Quest Pro controller was attached to the hanger suspending the conductive cloth. Inside Unity, the virtual cloth was parented to this tracked controller object, ensuring seamless spatial alignment and interaction between the physical and virtual environments, as shown in Figure~\ref{fig:setup}.


\subsubsection{Pinching Tactile Display}
\label{sec:tactile-feedback-system}

The PTD in our setup is centered on applying electrostatic forces to a conductive cloth. By varying voltage and frequency inputs, the system can adjust the roughness that users feel when pinching the cloth. We use a conductive cloth of approximately $25~\text{cm} \times 45~\text{cm}$, made of 100\% silver fiber. An Arduino UNO generates square waves at 50~Hz within 0--5~V. A MOS-FET (TK31N60X) switches these signals to deliver $100$--$300$~V to the cloth, effectively creating variable electrostatic forces. 
For safety measure, We used ultra-thin insulating polyvinyl chloride (PVC) gloves ($0.035~\text{mm}$ thick) to minimize direct exposure to high voltage.

\begin{figure}
 \centering
 \includegraphics[width=0.9\linewidth]{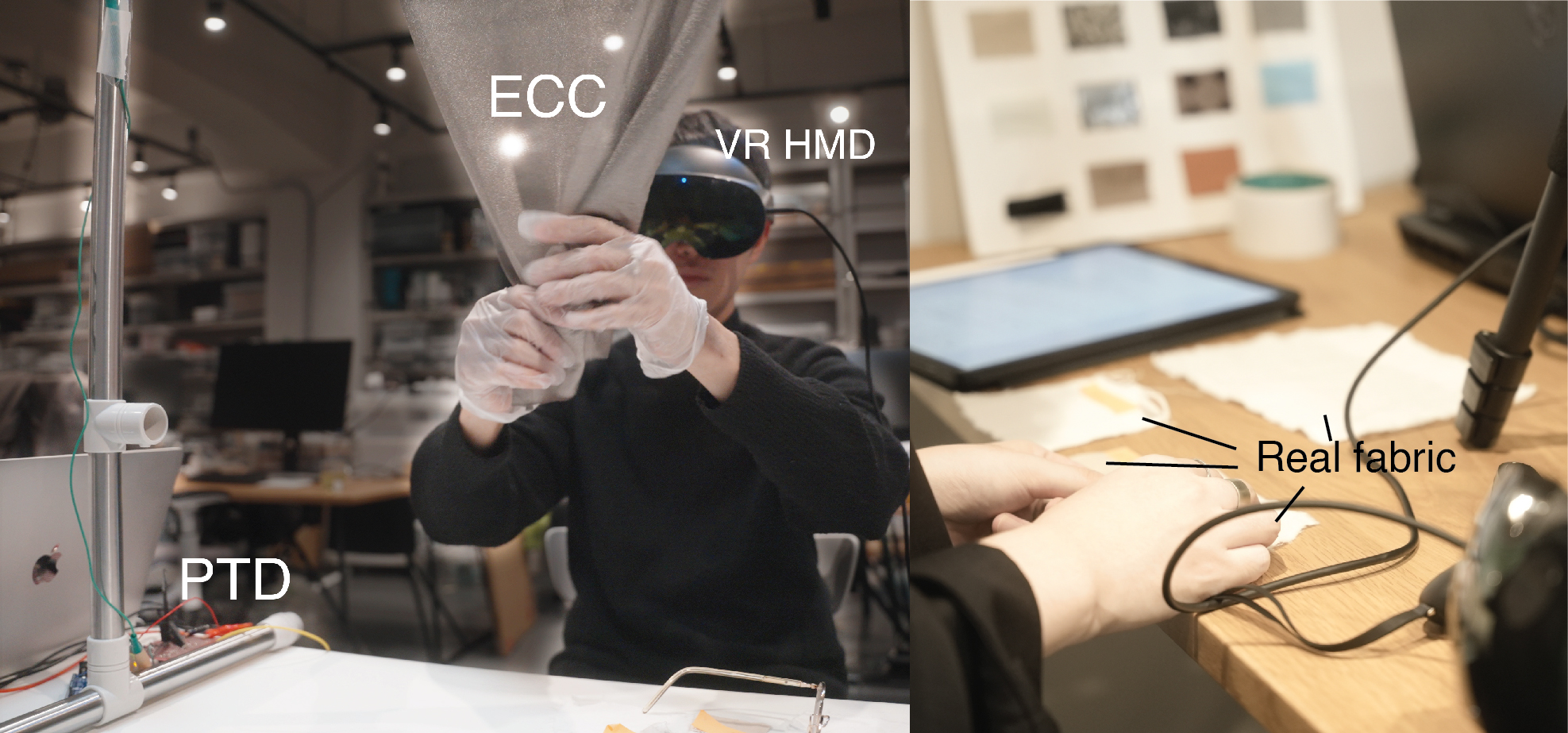}
 \caption{
 Left: shows how the participant interact with ECC in VR modality. Right: shows how the participant interact with real fabrics (denim, gauze, toweling, voile) as a baseline during the user experiment. 
 }  
 \Description[User study participant]{A participant wearing a VR headset touches a conductive cloth (ECC) with a PTD system on the left and real fabric samples on a wooden surface. This is for comparison testing in a laboratory setting.}
 \label{fig:experiment}
\end{figure}

\section{User Study}\label{sec:userstudy1}
To investigate how the visual texture overlay affects the tactile sensation provided by the PTD, we conducted a user study. Our focus was on the overall cross-modal effects of fabric interactions within a VR environment, particularly in the context of VR shopping.
Specifically, we examined how four different textures—denim, gauze, voile, and toweling—influenced the perception of fabric tactile properties. These textures were selected because they are commonly encountered in everyday items such as T-shirts, shirts, jeans, and towels.
Our goal is to determine whether the cross-modal effect extends the expressive capabilities of PTD beyond roughness to include other tactile properties, such as thickness, stiffness, and warmth.

\subsection{Experimental Design}
Our experiment used a 5×7 factorial design with two independent variables: visual conditions, voltage levels, and frequency settings. The visual conditions included either no VR headset, where participants directly viewed the physical PTD, or wearing a VR headset displaying one of four virtual fabric textures: denim, gauze, toweling, or voile. These textures were selected because they are commonly encountered in daily life. Seven distinct voltage levels were applied to the conductive fabric, specifically 0V, 50V, 100V, 150V, 200V, 250V, and 300V. The frequency was fixed at 50 Hz, as prior PTD studies showed it had no significant impact on the four fabric properties (roughness, stiffness, thickness, warmth). As dependent variables, participants rated four fundamental tactile properties of fabrics: roughness, stiffness, thickness, and warmth. These properties were assessed using a 5-point Likert scale. Our choice of descriptive term pairs (stiff-flexible, thick-thin, and soft-harsh) relates to the findings of Soufflet, Calonnier, and Dacremont~\cite{soufflet2004comparison} who demonstrate that the most significant scales their participants used to rate the properties of textiles were stiff-flexible, thick-thin, and soft-harsh. This design resulted in a total of 35 experimental conditions (7 voltage levels × 5 visual conditions) for each participant to evaluate. 

18 participants (8 males and 10 females, aged 19-54), balancing gender and age participated in the experiment. This sample size met G-Power analysis requirements. Participants were screened to ensure they were not experiencing fatigue or stress that might affect perception. The study protocol was approved by our Ethics Committee, and all participants provided written informed consent prior to participation.



\begin{table}[t]
    \centering
    \caption{Mapping of physical fabric samples to their corresponding positions on four perceptual dimensions (roughness, stiffness, thickness, and warmth), with each dimension represented as a bipolar scale.}
    \Description[Perceptual mapping of fabric samples]{Fabric positioning on four bipolar scales:
    Roughness: Jeans > Voile > Gauze
    Stiffness: Towelling > Gauze > Voile
    Thickness: Jeans > Towelling > Gauze
    Warmth: Towelling > Jeans > Voile}
    \label{tab:cloth_criteria}
    
    \begin{tikzpicture}[scale=0.8]
      \draw[<->] (0,0) -- (6,0);
      \draw (1.0,0.15) -- (1.0,-0.15);
      \draw (3.0,0.15) -- (3.0,-0.15);
      \draw (5.0,0.15) -- (5.0,-0.15);
      \node[left] at (0,0) {Rough};
      \node[right] at (6,0) {Smooth};
      \node[below] at (1.0,-0.15) {Jeans};
      \node[below] at (3.0,-0.15) {Voile};
      \node[below] at (5.0,-0.15) {Gauze};
      
      \draw[<->] (0,-1.5) -- (6,-1.5);
      \draw (1.0,-1.5+0.15) -- (1.0,-1.5-0.15);
      \draw (3.0,-1.5+0.15) -- (3.0,-1.5-0.15);
      \draw (5.0,-1.5+0.15) -- (5.0,-1.5-0.15);
      \node[left] at (0,-1.5) {Stiff};
      \node[right] at (6,-1.5) {Flexible};
      \node[below] at (1.0,-1.5-0.15) {Towelling};
      \node[below] at (3.0,-1.5-0.15) {Gauze};
      \node[below] at (5.0,-1.5-0.15) {Voile};
      
      \draw[<->] (0,-3.0) -- (6,-3.0);
      \draw (1.0,-3.0+0.15) -- (1.0,-3.0-0.15);
      \draw (3.0,-3.0+0.15) -- (3.0,-3.0-0.15);
      \draw (5.0,-3.0+0.15) -- (5.0,-3.0-0.15);
      \node[left] at (0,-3.0) {Thick};
      \node[right] at (6,-3.0) {Thin};
      \node[below] at (1.0,-3.0-0.15) {Jeans};
      \node[below] at (3.0,-3.0-0.15) {Towelling};
      \node[below] at (5.0,-3.0-0.15) {Gauze};
      
      \draw[<->] (0,-4.5) -- (6,-4.5);
      \draw (1.0,-4.5+0.15) -- (1.0,-4.5-0.15);
      \draw (3.0,-4.5+0.15) -- (3.0,-4.5-0.15);
      \draw (5.0,-4.5+0.15) -- (5.0,-4.5-0.15);
      \node[left] at (0,-4.5) {Warm};
      \node[right] at (6,-4.5) {Cool};
      \node[below] at (1.0,-4.5-0.15) {Towelling};
      \node[below] at (3.0,-4.5-0.15) {Jeans};
      \node[below] at (5.0,-4.5-0.15) {Voile};
    \end{tikzpicture}
    
\end{table}

\subsection{Procedure}

Figure~\ref{fig:experiment} presents the scene of the user experiment.
Participants interacted with the cloth under 35 different conditions, combining 7 types of voltage (0V, 50V, 100V, 150V, 200V, 250V, 300V) and 5 visual texture types (no texture, denim, gauze, toweling, voile). For conditions involving visual textures, participants wore a VR HMD and viewed the virtual cloth, rendered with the specified texture in Unity, which was spatially aligned with the physical conductive cloth. In the "no texture" condition, participants directly viewed the physical PTD without wearing the VR HMD.

For each condition, participants answered four questions about the tactile sensations of the cloth—roughness, thickness, stiffness, and warmth—on a 5-point Likert scale. To reduce the influence of individual differences in responses, we provided four reference fabric samples: jeans, voile, gauze, and toweling. These reference materials, classified according to tactile properties as shown in Table~\ref{tab:cloth_criteria}, represented distinct points along each of the four perceptual dimensions~\cite{atkinson2013tactile}. Participants could touch these samples at any time to use them as benchmarks for their evaluations. They compared each criterion by touching both the reference fabric and the conductive fabric to assign the most appropriate score. To eliminate the influence of the order in which the fabrics were touched and the tactile impression of the previously touched fabric, the order in which the conditions were presented was randomized. Each experimental session lasted approximately 90 minutes per participant.


\subsection{Data Analysis}
The data collected followed a within-subject two-factor design (voltage × texture). As normality could not be assumed for the data, we applied an aligned rank transform (ART) followed by two-way ANOVA~\cite{10.1145/1978942.1978963}. This method has been shown to be effective for non-parametric factorial analysis in HCI research. The ART procedure separates main effects and interactions by aligning the data for each effect before applying ranks, allowing the use of ANOVA procedures on the ranked data. This approach is particularly appropriate for our Likert scale data, which cannot be assumed to follow a normal distribution. We adjusted p-value limits using the Hommel method~\cite{hommel1988simultaneous} for multiple simultaneous comparisons.

\subsection{Results}
Table~\ref{tab:vr-texture-tactile-effects} shows the results. The results revealed differential effects of visual texture conditions on the dimensions of tactile perception. The results indicate that voile texture significantly affected stiffness ($p=.00887$); toweling texture tended to affect warmth ($p=.0658$); and denim texture tended to affect roughness ($p=.0972$). These results support the hypothesis from previous work that virtual texture overlay can affect the warmth, stiffness, and roughness of the PTD.  Since the PTD can only change roughness, these results suggest that texture overlay enables the display of tactile sensations that the PTD alone cannot express.  Additionally, we observed a marginally significant interaction between voile texture and voltage on stiffness ($p=.0536$).  This suggests that, while voile texture influences stiffness, its effect is dependent on voltage levels. On the other hand, contrary to our hypothesis, no significant influence of texture on thickness was observed.  We observed that most participants pinched the cloth. This tactile thickness may have led them to perceive it as reliable, causing them to underrate the visual information and preventing it from reaching a significant level.

    \begin{table}[t]
            \caption{P-values from ANOVA analysis following aligned rank transformation, examining the influence of four visual texture conditions on perceived tactile properties. Double asterisk (**) denotes significance at $p \leq .05$, and asterisk (*) denotes significance at $.05 <p \leq .1$.}
            \centering
            \begin{tabular}{lcccc}
            \toprule
            \textbf{Texture} & \textbf{Roughness} & \textbf{Stiffness} & \textbf{Thickness} & \textbf{Warmth} \\
            \midrule
             denim & .0972* & .704 & .882 & .446  \\
             gauze & .442 & .688 & .705 & .810 \\
             towelling & .270 & .574 & .686 & .0658*    \\
             voile & .148 & .00887** & .204 & .665  \\
            \bottomrule
            \end{tabular}
            \Description[Statistical results for texture effects]{P-values showing voile significantly affected stiffness (p=0.00857**), denim marginally affected roughness (p=0.0972*), towelling marginally affected warmth (p=0.0658*). *p≤0.1, **p≤0.05.}
            \label{tab:vr-texture-tactile-effects}
    \end{table}

\section{Discussion}\label{sec:discussion}
\subsection{Cross-Modal Integration in Tactile Perception}
Our findings demonstrate that visual textures significantly modulate fabric tactile perception, thereby expanding PTD's expressive capabilities beyond its inherent roughness control. This occurs differentially across properties.

For roughness, textures with clearly visible weave patterns (e.g., denim) significantly enhanced perception, correlating with their visual irregularity. This additive effect suggests that visual influences on roughness may remain consistent across different physical roughness levels, potentially enabling enhanced perception at lower voltage settings for safer haptic systems.

For stiffness and warmth, our results extend initial observations by showing significant visual influence. Specifically, voile texture significantly affected stiffness, and toweling texture tended to affect warmth. Since only roughness showed a significant change with PTD alone, this suggests that using the cross-modal effect enables the presentation of tactile sensations that could not be expressed by the conductive fabric alone. 

Conversely, for thickness, visual cues showed no significant influence, contradicting our hypothesis. This suggests that the strong tactile feedback from pinching the cloth may lead users to perceive thickness as highly reliable, causing them to underrate visual information. This aligns with theories of modality prioritization where the perceptual system emphasizes the most reliable sensory input for a given attribute, especially for spatial properties like thickness, making it less susceptible to visual deception.

\subsection{Limitations}
A key limitation of our study is the challenge of aligning deformation between real cloth and cloth in VR space.  Physical simulation made it difficult to perfectly match cloth deformation when touched.  Though the size and spatial position of the virtual cloth are 1:1 with the physical cloth, we acknowledge that perfectly aligning deformations, such as wrinkles, in real time between the physical and virtual cloth remains challenging. We also agree that quantitative alignment evaluation is crucial for future work.  Developing tracking systems using markers could address this issue, enabling more accurate reflection of physical cloth deformation in VR space and improving visual-tactile coherence.  

For texture selection and visual elements, a deeper understanding requires systematically investigating individual elements, such as fabric visual patterns (e.g., spatial frequency, contrast, or pattern regularity), or specific visual cues (e.g., dynamic shading, deformation animation, or temporal coherence).

\subsection{Future Applications and Directions}
Our findings suggest promising applications for enhancing PTD capabilities. Visual information can overcome the challenge of generating diverse tactile sensations from a single physical material.
Future work should thoroughly investigate the relationship between visual texture parameters (size, height, density of weave patterns) and roughness perception, including quantifying effects to provide design guidelines. Exploring combinations of different waveform patterns (sawtooth, sine waves) with visual textures could enable wider tactile expression.

Our findings also have important implications for improving tactile experiences in VR/AR environments. Future work should explore applications to more complex objects (e.g., clothing, sofas) and investigate visual-tactile integration effects on body parts beyond hands for full-body VR experiences. Developing projection mapping systems could enable testing visual-tactile integration in more natural conditions without VR headsets, leading to seamless mixed reality experiences.

\subsection{Safe and Responsible Innovation Statement}
Our visuo-haptic display aims to simulate fabric tactile sensations, enabling customers to feel clothing textures before purchasing online. This technology addresses significant social challenges in e-commerce by allowing consumers to confirm fabric quality by touch, significantly enhancing the online shopping experience. From an ethical perspective, our system contributes to reducing return rates and increasing customer satisfaction, thereby decreasing clothing waste—a major environmental concern in the fashion industry.

\section{Conclusion}\label{sec:conclusion}
In this paper, we developed the Pinching Visuo-Haptic Display to investigate how visual textures influence tactile perceptions of cloth. Our system aligns virtual cloth with physical electrostatic conductive cloth (ECC) for natural interaction in VR. Through user studies, we discovered that visual texture significantly modulates stiffness and shows tendencies for warmth and roughness, demonstrating how visual overlay expands PTD's expressive capabilities beyond its inherent roughness control. We also found thickness perception remained robust against visual influence. These findings offer insights for employing visual textures to enhance perceived fabric properties, paving the way for deployment in virtual fashion retail and VR fitting applications.

\begin{acks}
We recruited participants at https://www.jikken-baito.com for user study.
This work was supported by JST Moonshot R\&D Grant JPMJMS2012, JPNP23025 commissioned by the New Energy and Industrial Technology Development Organization (NEDO).
\end{acks}

\bibliographystyle{ACM-Reference-Format}
\bibliography{bib-base}

\appendix

\end{document}